\documentstyle[12pt]{article}
\begin{document}

\begin{center}

{\bf Numerical and approximate analytical results }

{\bf for the} {\bf frustrated spin}${\bf -}{\frac 12}$ {\bf quantum spin
chain.}

\bigskip
{R. Bursill}$^1${, G.A. Gehring}$^1${, D.J.J. Farnell}$^2${, J.B. Parkinson}$%
^2${,}

{\ } {\ Tao Xiang$^{1+}$ and Chen Zeng}$^2$

\bigskip\ $^1${\it Department of Physics, University of Sheffield, Sheffield}
S3 7RH.

\medskip\ $^2${\it Department of Mathematics, UMIST,} {\it P.O.Box 88,
Manchester} M60 1QD{\it .}
\end{center}

\bigskip

\medskip\ \noindent \underline{Abstract} We study the $T=0$ frustrated
phase of the $1D$ quantum spin-$\frac 12$ system with nearest-neighbour and
next-nearest-neighbour isotropic exchange known as the Majumdar-Ghosh
Hamiltonian. We first apply the coupled-cluster method of quantum many-body
theory based on a spiral model state to obtain the ground state energy and
the pitch angle. These results are compared with accurate numerical results
using the density matrix renormalisation group method, which also gives the
correlation functions. We also investigate the periodicity of the phase
using the Marshall sign criterion. We discuss particularly the behaviour
close to the phase transitions at each end of the frustrated phase.

\medskip\medskip

\noindent {\it +} Present address:

\noindent IRC in Superconductivity, University of Cambridge, Madingley Rd.,
Cambridge CB3 0HE.


\medskip
\noindent \underline{Short Title} Frustrated spin 1/2 chain.

\newpage\

\section{Introduction}
\setcounter{equation}{0}

There is currently much interest in quantum spin systems which exhibit
frustration. This has been stimulated in particular by the work on the
magnetic properties of the cuprates which become high T$_c$ superconductors
when doped. The frustration in these 2D materials arises because of
antiferromagnetic exchange across the diagonals of the squares as well as
along the edges. Other 2D frustrated systems are the triangular and Kagom\'e
lattices.

In this paper we study a simple 1D spin system which is also frustrated for
some range of its parameters. This is a spin-1/2 model with isotropic
nearest and next-nearest neighbour exchange given by

\begin{equation}
{\cal H}=\cos \omega \sum_l{\sl s}_l.{\sl s}_{l+1}+\sin \omega \sum_l%
{\sl s}_l.{\sl s}_{l+2},  
\end{equation}

\noindent where the sum over $l$ is over all $N$ atoms with periodic
boundary conditions. We shall also use the notation
$J_1=\cos \omega $
and $J_2=\sin \omega $.

The $T=0$ phase diagram of this model is given in Fig. 1. The
antiferromagnetic (AF) phase extends over the region
$-\pi /2<\omega<\omega _{MG}$, where $\omega _{MG}=\tan ^{-1}(1/2)$. The point
$\omega
_{MG}$ is the Majumdar-Ghosh (MG) Hamiltonian (Majumdar and Ghosh 1969a,b.
See also Haldane 1982) at which the ground state consists of dimerised
singlets with a gap to the excited states. In a recent paper by two of the
present authors (Zeng and Parkinson 1995), a dimer variational wave function
was proposed which is exact at $\omega _{MG}$ and gives good results for a
large range around this point.

Much of the recent work on this system has focused on the transition from a
gapless `spin-liquid' state which is known exactly at $\omega =0$ to a
dimerised regime with a gap which is also known exactly at
$\omega =\omega_{MG}$. The transition occurs at
$J_2/J_1=\tan\omega_c=0.2411(1)$
($\omega_c=0.2366(1)$) (Okamoto and Nomura 1992). The same authors have also
studied the phase diagram in the vicinity of this transition in the
anisotropic version of this model (Nomura and Okamoto, 1993,1994).

The frustrated regime is given by
$\omega _{MG}<\omega <\omega _{FF}$,
where $\omega _{FF}=\tan ^{-1}(-1/4)=3.3865$ is the point at which a first
order transition to a ferromagnetic regime occurs. This was first studied
numerically by Tonegawa and Harada (1987) who found evidence of change in
the position of the peak of the correlation function as a function of $%
\omega $. Here we shall use a variety of methods to investigate the whole
of the frustrated regime, including $\omega >\pi /2$.

It will be useful to compare our results with those of the classical
Hamiltonian. In this regime the minimum classical energy is obtained by
forming a spiral with a pitch angle $\theta $ between neighbouring spins
where $\theta =\cos ^{-1}(-J_1/4J_2)$. The classical boundary with the AF
phase is at $\omega _C=\tan ^{-1}(1/4)=0.2450$. The real-space periodicity
thus increases monotonically from 2 at the $AF$ boundary to infinity at the
ferromagnetic boundary.

\section{The CCM formalism}
\setcounter{equation}{0}

In a recent paper (Farnell and Parkinson, 1994, referred to as I), the
coupled-cluster method (CCM) was applied by two of the present authors to
the antiferromagnetic (AF) phase. For a description of the CCM applied to
spin systems see Bishop {\it et al. }(1991) and the references in I. In the
AF phase the natural choice of a model state for the CCM is the N\'eel state
used in I.

\noindent For the frustrated regime, however, this model state is physically
unrealistic and the CCM based upon it gives poor results. One possible
choice is suggested by the fact that when $\omega =\pi /2$ we have
$J_1=0 $ and $J_2=1$, so the Hamiltonian (1.1) describes two uncoupled
antiferromagnetic Heisenberg chains. At this point a `double-N\'eel' model
state with a periodicity of 4 unit cells would be appropriate and would lead
to precisely the same results as for the single chain $(J_1=1,J_2=0)$
with suitable scaling factors. We did carry out CCM calculations based on
this model state and obtained reasonable results for a range of $\omega $
around $\pi /2$. These results will be described briefly later.

Another possible model state is suggested by the classical ground state in
this regime. For this reason we have performed CCM calculations based on a
spiral model state in which the pitch angle $\theta $ is taken as a
variational parameter. A necessary condition to perform CCM calculations is
the existence of a complete set of mutually commuting creation operators so
that an arbitrary state of the system can be constructed starting from the
model state. We obtain these as follows.

The spiral model state is taken to have all spins aligned in the $XZ$ plane
with the n'th spin making an angle $n\theta $ with the $Z$ axis. We then
introduce local axes such that each atom is in the quantum spin state $\mid
->$. We use the usual notation $\mid \pm >$ for the states with eigenvalues
of $s^z$ equal to $\pm {\frac 12}$. Using the local axes the Hamiltonian
(1.1) becomes

\[
{\cal H}=J_1/4\sum_i\{[\cos (\theta
)-1](s_i^{-}s_{i+1}^{-}+s_i^{+}s_{i+1}^{+})+[\cos (\theta
)+1](s_i^{-}s_{i+1}^{+}+s_i^{+}s_{i+1}^{-})
\]
\[
+2\sin (\theta )(s_i^{-}+s_i^{+})(s_{i+1}^z-s_{i-1}^z)+4\cos (\theta
)s_i^zs_{i+1}^z\}
\]
\[
+J_2/4\sum_i\{[\cos (2\theta
)-1](s_i^{-}s_{i+2}^{-}+s_i^{+}s_{i+2}^{+})+[\cos (2\theta
)+1](s_i^{-}s_{i+2}^{+}+s_i^{+}s_{i+2}^{-})
\]
\begin{equation}
+2\sin (2\theta )(s_i^{-}+s_i^{+})(s_{i+2}^z-s_{i-2}^z)+4\cos (2\theta
)s_i^zs_{i+2}^z\}.  
\end{equation}

This equation contains terms which have an odd number of spin-flips
multiplied by a coefficient $\sin (\theta )$ or $\sin (2\theta )$. By
symmetry the ground-state energy $E_g$ will be an even function of $\theta $%
, which suggests that these terms should not contribute to $E_g$. We have
confirmed explicitly that this is correct for the CCM approximation scheme
described in the following section, and for clarity we shall omit these
terms from ${\cal H}$ from now on.

\subsection{Approximation schemes.}
\setcounter{equation}{0}

We shall work with Pauli spin operators $\sigma _i^\alpha $, related to the
spin angular momentum operators in the usual way: $\sigma _i^\alpha \sigma
_i^{\pm }=s_i^{\pm }$. These definitions apply to all sites as there is
no partition into different sublattices in this scheme. The Hamiltonian of
Eq.(1.1) becomes

\[
{\cal H}=J_1/4\sum_i\{[\cos (\theta )-1]
(\sigma _i^{-}\sigma_{i+1}^{-}+\sigma _i^{+}\sigma _{i+1}^{+})+
[\cos (\theta )+1](\sigma_i^{-}\sigma _{i+1}^{+}
+\sigma _i^{+}\sigma _{i+1}^{-})
\]
\[
+\cos (\theta )\sigma _i^z\sigma _{i+1}^z\}+J_2/4\sum_i\{[\cos (2\theta
)-1](\sigma _i^{-}\sigma _{i+2}^{-}+\sigma _i^{+}\sigma _{i+2}^{+})
\]
\begin{equation}
+[cos(2\theta )+1](\sigma _i^{-}\sigma _{i+2}^{+}+
\sigma _i^{+}\sigma_{i+2}^{-})+
\cos (2\theta )\sigma _i^z\sigma _{i+2}^z\}  
\end{equation}

In the CCM the true ground state is written
\begin{equation}
\mid \Psi >=e^S\mid \Phi >.  
\end{equation}
\noindent The CCM correlation operator $S$ is constructed entirely out of
creation operators with respect to the model state, i.e. out of a sum of
terms containing all possible $C_I^{+}$, where $C_I^{+}$is a product of
creation operators from $\{\sigma _i^{+}\}$ consistent with the conserved
quantities. The Hamiltonian of Eq.(2.1) contains only terms which involve an
even number of spin flips. This means that all terms in $e^S$ and hence in $%
S $ should only involve even numbers of $\sigma ^{+}$ operators. Note that
this would not be true had the $\sin (\theta )$ and $\sin (2\theta )$ terms
not been neglected, and this point is considered further below.

We shall use the following approximation schemes, all of which were
described in I.

\noindent 1) Full SUB2. In this scheme $S$ includes all possible products of
two spin-flip operators:

\begin{equation}
S={\frac 12}\sum_i\sum_rb_r\sigma _i^{+}\sigma _{i+r}^{+},  
\end{equation}
\noindent where i runs over all $N$sites and $r$is a positive or negative
integer with $\mid r\mid \le N/2.$ By symmetry $b_{-r}=b_r$.

\noindent 2) SUB2-3. This is a subset of full SUB2 in which all $b_r$are set
to zero except $b_{\pm 1}$: and $b_{\pm 2}$

\begin{equation}
S=b_1\sum_i\sigma _i^{+}\sigma _{i+1}^{+}b_1+b_2\sum_i\sigma _i^{+}\sigma
_{i+2}^{+}  
\end{equation}
$\qquad $

Using the same notation as in I we calculate the similarity
transform with respect to $S$ of the spin operators. For example

\begin{equation}
\tilde \sigma _i^{+}=e^{-S}\sigma _i^{+}e^S.  
\end{equation}
\noindent Using these the transformed Hamiltonian $\tilde {{\cal H}}$ can be
obtained. Operating on the ground state Schr\"odinger equation
\begin{equation}
\tilde {{\cal H}}\mid \Psi >=E_g\mid \Psi >  
\end{equation}
with $<\Phi \mid $ then gives the following equation for the ground-state
energy per spin in either approximation as

\begin{equation}
E_g/N=J_1/4\{\cos (\theta )+(\cos (\theta )-1)b_1)\}+J_2/4\{\cos (2\theta
)+(\cos (2\theta )-1)b_2)\}  
\end{equation}

To find $b_1$ and $b_2$ we obtain a set of coupled non-linear equations for
the coefficients retained in each of the approximation schemes by operating
on Eq.(2.6) with $<\Phi \mid C_I$, where $C_I$ is the Hermitian conjugate
of one of the strings of creation operators (combinations of $\sigma _i^{+})$%
present in S.

Lastly in this section we note that if odd numbers of spin flips been
allowed there would be a term in $S$ of the form
\[
a\sum_i\sigma _i^{+}.
\]

\noindent We have performed calculations in the SUB2-3 approximation in
which the extra $\sin (\theta )$ and $\sin (2\theta )$ terms were retained
in the Hamiltonian. In this case $a=0$ is the only physically reasonable
solution, and the extra terms give zero contribution to the ground-state
energy.

\section{The Coupled Non-linear Equations.}
\setcounter{equation}{0}

Using the $S$ given by Eq.(2.4), we operate on Eq.(2.7) with $\Sigma
_i\sigma _i^{-}\sigma _{i+t}^{-}$. and obtain the full SUB2 equations.

\[
J_1\sum_\rho (1-\delta _{r,0})\{A_1\delta _{r,\rho }+B_1b_r+2[\cos (\theta
)+1]b_{r+\rho }+[\cos (\theta )-1]\sum_sb_{r+s+\rho }b_s
\]
\begin{equation}
+J_2\sum_\delta (1-\delta _{r,0})\{A_2\delta _{r,\delta }+B_2b_r+2[\cos
(2\theta )+1]b_{r+\rho }+[\cos (2\theta )-1]\sum_sb_{r+s+\delta }b_s=0
\end{equation}

\noindent where
\begin{equation}
A_1=[\cos (\theta )-1](1+2b_1^2)+4b_1\cos (\theta ),  
\end{equation}
\begin{equation}
A_2=[\cos (2\theta )-1](1+2b_2^2)+4b_2\cos (2\theta ),  
\end{equation}
\begin{equation}
B_1=-4\cos (\theta )+4[1-\cos (\theta )]b_1  
\end{equation}
\begin{equation}
B_2=-4\cos (2\theta )+4[1-\cos (2\theta )]b_2  
\end{equation}
\noindent with $\rho =\pm 1,\delta =\pm 2$ and s is any positive or
negative integer. The solution of Eq.(3.1) is given in section 4.

For the SUB2-3 approximation scheme Eq.(3.1) reduces to the pair of coupled
non-linear equations

\[
J_1\{[\cos (\theta )-1](1+2b_2^2-3b_1^2)-4b_1\cos (\theta )+2b_2[[\cos
(\theta )+1]\}
\]
\begin{equation}
+J_2\{[1-\cos (2\theta )]4b_1b_2-8b_1\cos (2\theta )+2b_1[[\cos (2\theta
)+1]\}=0  
\end{equation}
and
\[
J_1\{[1-\cos (\theta )]4b_1b_2-8b_2\cos (\theta )+2b_1[[\cos (\theta )+1]\}
\]
\begin{equation}
+J_2\{[\cos (2\theta )-1](1+2b_1^2-3b_2^2)-4b_2\cos (2\theta )\}=0.
\end{equation}
Eqs.(3.4,5) can be solved numerically and hence $E_g/N$ obtained in the
SUB2-3 approximation for a given $\theta $. Finally $\theta $ is varied to
find a minimum value for $E_g/N$.

The results for $\theta $ as a function of $\omega $ are shown in Fig. 2. We
observe that the value of $\theta $ obtained by this method remains close to
$\pi /2$ over a much wider range of $\omega $ than in the classical
calculation. We mentioned earlier that calculations based on a
`double-N\'eel' model state have been carried out. As can now be easily
understood, the results were in good agreement with the ones based on the
spiral model state over quite a wide range of $\omega $ around $\pi /2.$

The results for the ground-state energy per spin are shown in Fig.3 and are
compared with the values obtained by direct diagonalisation of a chain of 20
spins, the results of spin-wave theory (SWT), and also with a `classical'
result which is the expectation value of the Hamiltonian in the classical
ground-state. The exact results at $\omega =\omega _{MG}$ and $\omega =\pi $.

The full SUB2 equations can be solved numerically by first performing a
Fourier transform as in I. Details are given in Appendix 1. The results are
similar to the SUB2-3 results except for the existence of `terminating
points' which are also shown on the figures.

\section{DMRG study of the periodicity}
\setcounter{equation}{0}

We next turn to the density matrix renormalisation group (DMRG) method in
order to perform a numerical study of the periodicity which can be compared
with the CCM results discussed above. We achieve this by accurately
calculating the position of the peak of the Fourier transformed ground state
correlation function (Bursill {\em et al} 1995).

\subsection*{The DMRG method}

The DMRG was introduced in a series of papers by White and coworkers (White
and Noack 1992, White 1992 and 1993) and a highly successful application to
the spin-1 antiferromagnetic chain (White and Huse 1993) established the
DMRG as the method of choice for studying the low energy physics of quantum
lattice systems in one dimension. Efficient algorithms for calculating
low-lying energies and correlation functions of spin chains are described in
great detail in (White 1993) so we will only briefly describe the method
here. We restrict our discussion to the infinite lattice algorithm (White
1993) which was used in our calculations.

The DMRG is an iterative, truncated basis procedure whereby a large chain
(or superblock) is built up from a single site by adding a small number of
sites at a time. At each stage the superblock consists of system and
environment blocks (determined from previous iterations) in addition to a
small number of extra sites. Also determined from previous iterations are
the matrix elements of various operators such as the block Hamiltonians and
the spin operators for the sites (at the end(s) of the blocks) with respect
to a truncated basis. Tensor products of the states of the system block, the
environment block and the extra sites are then formed to provide a truncated
basis for the superblock. The ground state $\left| \psi \right\rangle $ (or
other targeted state) of the superblock is determined by a sparse matrix
diagonalization algorithm.

At this point, correlation functions, local energies and other expectation
values are calculated with respect to $\left|\psi\right>$. Next, a basis for
an augmented block, consisting of the system block and a specified choice of
the extra sites, is formed from tensor products of system block and site
states. The augmented block becomes the system block in the next iteration.
However, in order to keep the size of the superblock basis from growing, the
basis for the augmented block is truncated. We form a density matrix by
projecting $\left|\psi\right>\left<\psi\right|$ onto the augmented block
which we diagonalise with a dense matrix routine. We retain the {\em most
probable} eigenstates (those with the largest eigenvalues) of the density
matrix in order to form a truncated basis for the augmented block that is
around the same size as the system block basis. Matrix elements for the
Hamiltonian and active site operators, together with any other operators
that are required for say, correlation functions are then updated.

The environment block used for the next iteration is usually chosen to be a
reflected version of the system block. The initial system and environment
blocks are chosen to be single sites.

The accuracy and computer requirements of the scheme is fixed by $n_{%
\mbox{\protect\scriptsize s}}$, the number of states retained per block (of
good quantum numbers) at each iteration. $n_{\mbox{\protect\scriptsize s}}$
determines the truncation error, which is the sum of the eigenvalues of the
density matrix corresponding to states which are shed in the truncation
process. The error in quantities such as the ground state energy scale
linearly with the truncation error (White and Huse 1993).

\subsection*{Application of the DMRG to the frustrated spin-$1/2$ chain}

We have applied the infinite lattice DMRG algorithm to (1.1) using a number
of superblock configurations and boundary conditions. All the interactions
(intrablock, interblock and superblock Hamiltonians) commute with the total $%
z$ spin $S_{\mbox{\protect\scriptsize T}}^z\equiv \sum_iS_i^z$, so $S_{%
\mbox{\protect\scriptsize T}}^z$ is a good quantum number which can be used
to block diagonalize the system, environment and super blocks. For even
numbers of sites, the ground state of the superblock $\left| \psi
\right\rangle $ is a singlet with zero total spin so we only need to
consider superblock states with $S_{\mbox{\protect\scriptsize T}}^z=0$. We
found that the most cpu efficient configuration was the standard open ended
superblock of the form system-site-site-environment (White 1993).

As mentioned, in applying the DMRG to (1.1), we are concerned with the
correlation function
\begin{equation}
C_{jl}\equiv \left< S_{j}^{z}S_{l}^{z} \right>  
\end{equation}
and hence its Fourier transform
\begin{equation}
\tilde{C}(q)= \frac{1}{V}\sum_{jl}C_{jl}e^{iq(j-l)}  
\end{equation}
We are particularly interested in $q^{*}$, the value of $q$ where $\tilde{C}%
(q)$ has its peak. This leads to a natural (working) identification of the
ground state periodicity with $2\pi/q^{*}$ which was given in (Bursill {\em %
et al} 1995) where another frustrated spin model, the spin-1 model with
bilinear and biquadratic exchange, was studied.

In practice, $C_{jl}$ is calculated with $j$ and $l$ approximately
equidistant from the centre of the superblock and far from the ends of the
block so as to avoid end effects. In forming $\tilde C(q)$ we calculate $%
C_{jl}$ for $0\leq |j-l|\leq 60$. The algorithm is iterated until these
quantities converge. We test the algorithm by exactly calculating $C_{jl}$
for finite chains of up to 20 sites using the Lanzcos method and ensuring
that these results are reproduced by the DMRG.

In (Bursill {\em et al} 1995) it was noted that there are two impediments to
an accurate calculation of (4.2). Firstly, for given $j$ and $l$, we must
have $n_{\mbox{\protect\scriptsize s}}$ sufficiently large that $C_{jl}$ is
accurately determined. Secondly, for given $q$, we must retain enough
accurately calculated $C_{jl}$ in truncating the infinite series to
ensure an accurate result. It was found (Bursill {\em et al} 1995) that if
the system has a significant energy gap and exponentially decaying
correlation functions with a short correlation length, then the $C_{jl}$
converge rapidly with $n_{\mbox{\protect\scriptsize s}}$ and the Fourier
series converges very rapidly. On the other hand, in critical or near critical
regions where the energy gap is small or zero and the correlation functions
decay algebraically or have a large correlation length then convergence is
very slow.

By choosing $n_{\mbox{\protect\scriptsize s}}$ up to 90, it is found that
the main source of inaccuracy in calculating $\tilde{C}(q)$ in these regions
is Fourier series truncation. We plot $\tilde{C}(q)$ as a function of $q$
for various values of $J_{2}/J_{1}$ in Fig. 4.

As mentioned, it was determined using exact diagonalization and finite
size scaling methods (Okamoto and Nomura 1992) that the model is critical
(gapless with algebraically decaying correlation functions) for
$0\leq J_2/J_1\leq\tan\omega_{\mbox{\protect\scriptsize c}}$ and gapped
beyond this region where $\tan\omega_{\mbox{\protect\scriptsize c}}=0.2411(1)$
Correspondingly, we find that $\tilde{C}(q)$ converges slowly and has
oscillation due to Fourier series truncation in and around the critical
region. In the region $0.3\leq J_2/J_1\leq 2$ we find that $\tilde{C}(q)$
converges rapidly to a smooth function.

Now at the extreme point where $J_1=0$ we have two decoupled Heisenberg
chains and so $C_{jl}$ vanishes if $j$ and $l$ lie on different sublattices
but $C_{jl}$ decays algebraically on a given sublattice. We in fact find
that $\tilde{C}(q)$ converges slowly for $J_2/J_1\geq 2.5$ indicating that
there may be a finite interval around the $J_1=0$ point where the model is
critical.

We next turn to the question of periodicity in the ground state. As
mentioned, we define the periodicity in terms of the position $q^{*}$ at
which $\tilde C(q)$ has its peak. A plot of $q^{*}$ as a function of
$\omega$ is included in Fig. 2. We see that the simple, analytical CCM
result for the pitch angle improves dramatically upon the classical result.
Also, we see that the dimer variational wavefunction (Zeng and Parkinson
1995) gives an excellent estimate of the pitch angle in a region to the
right of the solvable point.

$q^{*}$ converges very rapidly with $n_{\mbox{\protect\scriptsize F}}$ (the
number of Fourier coefficients used in forming (4.2)) and
$n_{\mbox{\protect\scriptsize s}}$ in the region $0.3<J_2/J_1<2$ and we can
accurately determine the threshold (the onset of the spiral phase) $\tilde
\omega $ at which $q^{*}$ begins to move away from $\pi $ (as the
periodicity begins to change from 2 to 4). Such a threshold was found in
(Bursill {\em et al} 1995) as the biquadratic interaction was increased
relative to the bilinear interaction. Again $q^{*}$ could be accurately
determined near the threshold. Using the same analysis as in (Bursill {\em %
et al} 1995) then, we find
\begin{equation}
\tan \tilde \omega =0.52063(6)  
\end{equation}
This is to be compared with the classical threshold (0.25) and the
terminating point from the CCM theory (0.557).

In a recent preprint (Chitra {\it et al} 1994) have studied the extension of
(1.1) where there is also dimerization $\delta $ such that nearest neighbour
exchange carries a factor of $1+\delta $ and $1-\delta $ on successive
bonds. They conjectured that there is a disorder line given by $%
J_2/J_1=\frac 12(1-\delta )$ such that, in the $\delta $-$J_2/J_1$ plane,
the structure factor has its peak at $\pi $ below the line and decreases
from $\pi $ to $\pi /2$ as $J_2/J_1$ is increased above the line. In the
case (1.1) of no dimerisation $(\delta =0)$ gives $\tan $ $\tilde \omega
=1/2 $ (i.e. the threshold is the exactly solvable point).

Now at the solvable point the ground state is a perfect dimer where spins
form a singlet with their dimer pair but are otherwise uncorrelated. The
correlation function is

\[
C_{ij}=\left\{
\begin{array}{l}
\frac{1}{4}\mbox{ for }i=j \\
-\frac{1}{4}\mbox{ for }i\mbox{ and }j\mbox{ on the same dimer} \\
0\mbox{ otherwise}
\end{array}
\right.
\]

The Fourier transform is therefore
\begin{equation}
\tilde C(q)=\frac 14(1-\cos q)
\end{equation}
whence $\tilde C^{\prime \prime }(\pi )=-1/4\ne 0$ so, unless $\tilde
C^{\prime \prime }(\pi )$ is highly singular at the threshold, the threshold
(i.e. the point where $\tilde C^{\prime \prime }(\pi )$ vanishes) cannot
occur at the solvable point. This is borne out by our result (4.3).

\subsection*{Further interpretation of the spiral phase}

We have defined the ground state periodicity and the spiral phase ($\tilde w$%
) in terms of the peak position of the Fourier transformed correlation
function. It has, however been shown by (Schollw\"ock {\it et al} 1995) that
further insight into disorder and incommensurate spin distortions in the
ground state can be gained by investigating the correlation function in real
space. In Table 1 we list the correlation function in real space $C(r)$ for $%
J_2/J_1=$0.49, 0.5 (the solvable point), 0.51 and 0.5206\ldots (the
threshold). (As we shall see, in the gapped region, the ground state
has broken translational symmetry and $C(r)$ is defined to be the average
of $C_{j\;j+r} $ over a number of the sites $j$ in the middle of the chain).

We see that modulations begin to appear for $J_2/J_1$ values between the
solvable point and the threshold where $C(2)$ changes sign. That is, the
Majumdar-Gosh point {\it is} a disorder point, separating phases of
commensurate and incommensurate correlations (in real space). Following
(Schollw\"ock {\it et al} 1995), the threshold $\tilde \omega $, where
incommensurate spin oscillations would begin to be observed (experimentally)
in the structure factor, is identified as a Lipshitz point. We would expect
that in the limit of large spin $S$, the classical disorder point, the
quantum disorder point and the Lipshitz point would merge, there being a
single point separating commensurate and incommensurate phases both in terms
of real and momentum space.

\subsection*{Translational symmetry breaking in the ground state---a dimer
order parameter}

As mentioned, (Okamoto and Nomura 1992) calculated the critical point
$\tan\omega _{\mbox{\protect\scriptsize c}}=0.2411(1)$ separating gapped from
gapless phases. (Chitra {\it et al} 1994) calculated the energy gap using
the DMRG and deduced $\tan\omega _{\mbox{\protect\scriptsize c}}=0.298(1)$,
a result which is incompatible with that of (Okamoto and Nomura 1992). It is
however known (White 1993, Bursill {\em et al} 1995 and Schollw\"ock {\em et
al} 1995) that it is difficult to obtain accurate energies with the DMRG for
critical or near-critical systems. This is again borne out when we apply the
DMRG to the calculation of another order parameter which characterizes this
phase transition.

It is known that the ground state for the Heisenberg model $J_2=0$ has no
symmetry breaking whereas at the Majumdar-Gosh point $J_2=J_1/2$ the ground
state has broken translational symmetry, the correlator $C_{j\;j+1}$
equating to $0$ and $-1/4$ on successive bonds $(j,j+1)$. To measure this
broken symmetry, we define a dimer order parameter $D$ by
\[
D(N)\equiv \left| C_{N/2-1\;N/2}-C_{N/2\;N/2+1}\right| 
\]
and $D\equiv \lim_{N\rightarrow \infty }D(N)$ where $N$ is
the size of an even, open chain.

$D(N)$ converges very slowly in and around the critical region
$0\leq J_2/J_1<0.35$ and rapidly (with respect to both $N$ and
$n_{\mbox{\protect\scriptsize s}}$) around the threshold $0.45\leq J_2/J_1<1$.
A plot of $D$ versus $J_2/J_1$ for $n_{\mbox{\protect\scriptsize s}}=40$ is
given in Fig. 5. We note that $D$ is maximal at around $J_2/J_1\approx 0.58$
i.e. neither the disorder point (0.5) nor the Lipshitz point (0.52\ldots ).
The fact that $D$ exceeds 1/4 to the right of the disorder point is
indicative of the incommensurate oscillations whereby the values of
$C_{j\;j+1}$ on successive bonds $(j,j+1)$ can have opposite sign. We see
that the critical point is not well defined and only qualitative information
about the phase transition can be deduced from this procedure. We shall
attempt to address the question of how the DMRG can be adapted to study
critical phenomena in future publications.

\section{The Marshall sign results.}
\setcounter{equation}{0}

An additional method of studying the periodicity of the ground state in the
frustrated phase is by means of the Marshall-Peierls (Marshall, 1955) sign
criterion. Preliminary results were reported in an earlier paper (Zeng and
Parkinson, 1995) so a detailed description will not be given here. We have
now obtained results for an open chain of 16 atoms and these confirm and
extend those of shorter chains.

In Fig. 6 we show the parameter $\rho _i$ for $i=1,2,3,4$, corresponding
to a periodicity of $2i$ in the 16 atom chain. This parameter will be close
to $1$ if the ground state `conforms' to the given periodicity and will be
close to $0.5$ if the conformity is poor. The main features are as follows.

For $\omega $ in the range $0\leq \omega \leq \omega _{MG}$ (outside the
frustrated regime) $\rho _1$ is very close to $1$. For $\omega _{MG}\leq
\omega $ there is an extended region in which $\rho _2$ is closest to $1$.
An interesting and totally unexplained feature is the shallow double minimum
in the value of $\rho _2$ for $\omega $ near $\pi /4$, which was also
observed for shorter chains. At $\omega \approx 2.74$ there is a smooth
crossover to a state in which $\rho _3$ is largest and finally a more
complicated behaviour as $\omega $ approaches $\omega _{FF}$. An enlarged
picture of the latter region is shown in Fig. 7. The sharp changes in $\rho
_3$ at $\omega \approx 2.82$ and $2.85$ are caused by the crossing of a
quintuplet state to become the ground state between these two values. This
may be a `small N' effect, although even here $\rho _3$ is larger than the
other $\rho _i$. Finally we observe a region closer to $\omega _{FF}$ in
which $\rho _4$ is the largest. Results in this area are difficult to obtain
because there are many states lying close to the ground state and
convergence is extremely slow.

Nevertheless, these results do suggest that the periodicity in the
frustrated regime increases as the ferromagnetic boundary is approached. At
present, the quantum system looks rather different to the classical as the
change in periodicity occurs as a sequence of crossovers rather than
smoothly. However the chains are still relatively short and it may well be
that in the large $N$ limit the behaviour would approximate more closely to
the classical.

\section{Conclusion}
\setcounter{equation}{0}

The quantum mechanical behaviour of the frustrated phase of this system is
clearly rather complex. The picture that is beginning to emerge is that the
variation in periodicity with $\omega $ that is characteristic of the
classical ground state may well survive partially in the quantum system.
However, there are clearly many differences in detail and also some
completely new features.

The main difference in detail is that the periodicity of the quantum system,
as predicted by the Coupled-cluster method and the variational method and
confirmed by the DMRG results, remains closer to $\pi /2$ over a much wider
range of $\omega $ than does the classical system. Another difference,
suggested by the Marshall sign calculations, is that the changes in
periodicity close to the ferromagnetic boundary may occur less smoothly.

The behaviour of the quantum system close to the Majumdar-Ghosh point is
quite different, as there is no classical analogue of the highly dimerised
nature of the ground state.

\appendix
\section{Appendix 1. Solution of the full SUB2 equations.}
\setcounter{equation}{0}

The full SUB2 equations, Eq.(3.1), can be solved using Fourier transforms as
described in Appendix A of I. The result is

\begin{equation}
b_r={\frac 1\pi }\int_0^\pi dq\cos (rq)\Gamma (q)  
\end{equation}
\noindent with $\Gamma (q)$ given by

\begin{equation}
\alpha \Gamma (q)^2+\beta \Gamma (q)+\gamma =0  
\end{equation}
\noindent where

\[
\alpha =J_1\cos (k)[\cos (\theta )-1]+J_2\cos (2k)[\cos (2\theta )-1]
\]

\[
\beta =J_1\{B_1+2\cos (k)[\cos (\theta )+1]\}+J_2\{B_2+\cos (2k)[\cos
(2\theta )+1]\}
\]

\[
\gamma =J_1\{A_1\cos (k)-2b_1[\cos (\theta )+1]+[1-\cos (\theta )]X_1\}
\]
\[
+J_2\{A_2\cos (2k)-2b_2[\cos (2\theta )+1]+[1-\cos (2\theta )]X_2\}
\]
\noindent where
\[
X_1=\sum_sb_sb_{s+1};\qquad X_2=\sum_sb_sb_{s+2}.
\]

These equations are then solved numerically by constructing self-consistency
equations in the coefficients $b_1$ and $b_2$ and also $X_1$ and $X_2$.
Again $\theta $ is varied to find the minimum $E_g$.

\noindent {\bf \ Acknowledgements}

We have benefited greatly from discussions with R.F. Bishop and Yang Xian.
D.J.J. Farnell acknowledges a postgraduate award from the Science and
Engineering Research Council of Great Britain. R. Bursill is supported by
SERC, under grant GR/J26748.

\newpage\

\noindent {\bf References}

\noindent R.F.Bishop, J.B.Parkinson and Yang Xian, 1991, Phys.Rev. B {\bf 44}%
, 9425-43

\noindent R.J.Bursill, T.Xiang and G.A.Gehring, 1995, J.\ Phys. A

\noindent R.Chitra, Swapan K.Pati, H.R.Krishnamurthy, D.Sen and S.Ramasesha,
1994, {\em DMRG Studies of the Spin-Half Heisenberg System with Dimerization
and Frustration} Preprint

\noindent D.J.J.Farnell and J.B.Parkinson, 1994, J.Phys.:CM {\bf 6},
5521-5532

\noindent F.D.M.Haldane, 1982, Phys. Rev.B {\bf 25}, 4925-8

\noindent K.Majumdar and D.K.Ghosh, 1969a, J.Math. Phys. {\bf 10}, 1388-98

\noindent K.Majumdar and D.K.Ghosh, 1969b, J.Math. Phys. {\bf 10}, 1399-1402

\noindent W.Marshall, 1955, Proc. Roy. Soc. A {\bf 232,} 48-68

\noindent K.Nomura and K.Okamoto, 1993, J.Phys. Soc. Japan {\bf 62}, 1123-6

\noindent K.Nomura and K.Okamoto, 1994, J.Phys. A: Math. Gen. {\bf 27,}
5773-88

\noindent K.Okamoto and K.Nomura, 1992, Phys. Lett. A {\bf 169,} 433-7

\noindent U.Schollw\"ock, Th.Jolicoeur and T.Garel, 1995, {\em Physical
meaning of the Affleck-Kennedy-Lieb-Tasaki $S=1$ quantum spin chain} Preprint

\noindent T.Tonegawa and I.Harada, 1987, J. Phys. Soc. Japan {\bf 56},
2153-67

\noindent S.R.White and R.Noack 1992, Phys. Rev. Lett. {\bf 68} 3487-

\noindent S.R.White, 1992, Phys. Rev. Lett. {\bf 69} 2863-6

\noindent S.R.White, 1993, Phys. Rev. B {\bf 48} 10 345-

\noindent S.R.White and D.A.Huse, 1993, Phys. Rev. B {\bf 48} 3844-

\noindent Chen Zeng and J.B.Parkinson, 1995, Phys. Rev. B, {\bf 51 }11609-

\newpage
\noindent \underline{Figure Captions}.

\medskip\ \noindent \underline{Figure 1} $T=0$ phase diagram of the model.

\medskip\ \noindent \underline{Figure 2} Pitch angle $\theta $ as a function
of $\omega $ obtained by various methods.

\medskip\ \noindent \underline{Figure 3} Ground-state energy per spin as a
function of $\omega $. Open circles are exact results. Closed circle is the
CCM terminating point.

\medskip\ \noindent \underline{Figure 4} Fourier transformed correlation
functions for various values of $J_2/J_1$ obtained from the DMRG.

\medskip\ \noindent \underline{Figure 5} Dimer order parameter $D$ as a
function of $J_2/J_1$ obtained from the DMRG.

\medskip\ \noindent \underline{Figure 6} Marshall sign parameters $\rho _i$
as a function of $\omega $.

\medskip\ \noindent \underline{Figure 7} Enlarged part of Fig. 5 showing
region close to $\omega _2$.

\newpage

\noindent \underline{Table Caption}.

\medskip\ \noindent \underline{Table 1} Correlation function in real space
$C(r)$ for various values of $J_2/J_1$ obtained from the DMRG.

\newpage\
\noindent \underline{Table 1}.

\begin{tabular}{ccccc}
$r$ & 0.49 & 0.5 & 0.51 & $\tan\tilde{\omega}$ \\ \hline
0 & 0.25 & 0.25 & 0.25 & 0.25 \\
1 & -0.127 & -0.125 & -0.123 & -0.121 \\
2 & 0.00386 & 0 & -0.0039 & -0.00806 \\
3 & -0.00237 & 0 & 0.00234 & 0.00477 \\
4 & 0.000892 & 0 & -0.000764 & -0.00143 \\
5 & -0.000571 & 0 & 0.000425 & 0.000714 \\
6 & 0.00022 & 0 & -0.000119 & -0.000145
\end{tabular}

\end{document}